\documentstyle[12pt]{article}
\begin{document}

\begin{center}
{\large \bf Nature of traps responsible for failure of MOS devices}
\end{center}

\begin{center}
{V.A. Gritsenko$^*$, P.M. Lenahan$^a$, Yu.N. Morokov$^{b}$, and Yu.N. Novikov}
\end{center}

{Institute of Semiconductor Physics, Novosibirsk, 630090, Russia \\
$^a$ The Pennsylvanian State University, University Park, Pennsylvania 16802 \\
$^{b}$ Institute of Computational Technologies, Novosibirsk, 630090, Russia}
 
\begin{abstract}
{A failure of chips in a huge amount of modern electronic devices such as 
computers, TV, radio etc is connected as a rule with the undesirable capturing 
of charge (electrons and holes) by traps in thin insulating film of silicon 
oxide (SiO$_2$) in transistors. It leads to a breakdown of transistors or to a 
destructive change of their characteristics. It is suggested that silicon 
oxide will be replaced in the next generation of nanoscale devices by silicon 
oxynitride (SiO$_x$N$_y$) [1]. Therefore, it is very important to understand 
the nature of traps in this material. We discuss this nature using the 
quantum-chemical simulation.}
\end{abstract}

\vspace{5mm}

More than 90 percent of electronic devices are based on silicon. More than 
85\% of silicon devices are based on Metal-Oxide-Semiconductor (MOS) 
transistors. The MOS transistor consists of silicon substrate, source, drain, 
and gate (Fig. 1). Gate is separated from silicon by SiO$_2$. The traps in the 
thin film of SiO$_2$ can capture electrons or holes from a channel. This is 
one of the main source of the failure of electronic devices. A key unsolved 
problem with SiO$_2$ as a gate dielectric is a hole capture in the Si-Si 
defects near the Si/SiO$_2$ interface and the following device degradation 
[2]. 

Silicon oxide, which has been used as the gate dielectric more than thirty 
years, will be replaced in next generation of nanoscale MOS devices by silicon 
oxynitride [1]. A gate oxynitride fabrication with high-temperature annealing 
of thermal SiO$_2$ in ammonia results in a hole traps removal from the 
Si/SiO$_2$ interface [3] (Fig. 2). However, such oxynitride contains 
undesirable large density of electron traps. To improve and to control the 
dielectric properties it is important to understand which structural defects 
are the dominant traps in oxide and oxynitride. In the present paper we 
consider, using the quantum-chemical simulation, the nature of such traps in 
SiO$_x$N$_y$ and mechanism of creation and removal of traps. 

The defects and traps are absent in "ideal" oxynitride containing only the 
Si-O and Si-N bonds. The oxygen and nitrogen atoms in amorphous silicon 
oxynitride as in SiO$_2$ and Si$_3$N$_4$ (silicon nitride) are bonded with two 
and three silicon atoms, respectively [4]. The Si atoms in these materials are 
bonded with four other atoms (O or N). The traps in SiO$_x$N$_y$ are related 
with defects in atomic network. 

The main defects, identified in SiO$_x$N$_y$, are the next: the 
silicon-silicon bond ($\equiv$Si-Si$\equiv$), the three-fold coordinated 
silicon atom $\equiv$Si*, the two-fold coordinated nitrogen atom 
$\equiv$Si$_2$N*, and the non-bridged oxygen atom $\equiv$SiO*. Here "*" marks 
one unpaired electron. The main defects related with impurities are the 
hydrogen bonds: $\equiv$SiOH, $\equiv$Si$_2$NH, and $\equiv$SiH. 

To understand the nature of traps in SiO$_x$N$_y$ we considered the capturing 
properties of the main defects in SiO$_x$N$_y$. We have studied the electronic 
structure of the clusters (small fragments of material) in different charge 
states corresponding to the capture of electrons or holes by traps. The 
dangling bonds at the cluster surface were saturated with hydrogen atoms. 
Atomic relaxation of the atoms in the first coordination sphere of defects was 
considered in the simulation. The calculations were maid with the 
quantum-chemical method MINDO/3 with the parameters used in Refs. [5,6]. 

We used the Si$_{18}$N$_{17}$H$_{35}$ and Si$_{20}$N$_{28}$H$_{36}$ clusters 
for simulation of the Si$_3$N$_4$ bulk electronic structure. The clusters with 
different numbers of oxygen and nitrogen atoms in the second coordination 
sphere were considered to study the effect of the chemical composition on the 
capturing properties of the defects in SiO$_x$N$_y$. We estimated the energy 
gain for the electron or hole capture from the bulk of material as the 
differences between the total energies of the bulk and defect clusters in 
different charge states. 

Simulation of the neutral $\equiv$Si$_2$N* defect (nitrogen dangling bond) [6] 
shows that unpaired electron is localized for all considered clusters in a 
nitrogen 2p$_{\pi}$ orbital, which is oriented normally to the Si$_2$N plane. 
This result agrees with the previous theoretical simulation of this defect 
[7,8] and with the ESR experiments [9,10] for the $\equiv$Si$_2$N* defect in 
Si$_3$N$_4$. The positive energy gain for electron capture by the 
$\equiv$Si$_2$N* defect in Si$_3$N$_4$ is about 0.8 eV. It means that the 
defect can capture an electron. The calculated results show that the 
$\equiv$Si$_2$N* defect in silicon nitride and in oxynitride with a high 
concentration of nitrogen can capture an electron and cannot capture a hole 
[6]. 

The analogue of the $\equiv$Si$_2$N* defect in Si$_3$N$_4$ is the $\equiv$SiO* 
defect in SiO$_2$. It was shown in Ref. [11] that the $\equiv$SiO* defect is 
the electron trap in SiO$_2$. Our calculations for the $\equiv$SiO* defect in 
SiO$_x$N$_y$ and SiO$_2$ confirm this result. Thus, the obtained results 
together with experiments show that the $\equiv$Si$_2$N* and $\equiv$SiO* 
defects are the electron traps in gate dielectric. 

The $\equiv$Si$_2$N* defect creation in SiO$_x$N$_y$ is related with a 
breaking of the $\equiv$Si$_2$N-H bond according to the reaction [10]
\begin{center}
           $\equiv$Si$_2$NH $\rightarrow$ $\equiv$Si$_2$N* + H.                 
\end{center}
We associate the electron traps in oxynitride with these $\equiv$ Si$_2$N* 
defects. In order to remove the electron traps from gate oxynitride in silicon 
devices it is used the reoxidation of oxynitride. The removal of the electron 
traps at oxynitride reoxidation is explained by replacing of the =N* atom with 
oxygen atom according to the reaction 
\begin{center}
           $\equiv$Si$_2$N* + O $\rightarrow$ $\equiv$Si-O-Si$\equiv$ + N.  
\end{center}
With this reaction we can understand the chemical nature of one of the key 
technological step in the MOSFET device fabrication - the electron traps 
removal at the gate silicon oxynitride reoxidation (Fig. 2). 

The calculations have shown that the main extrinsic defects, related with the 
hydrogen bonds, $\equiv$SiOH, $\equiv$Si$_2$NH, and $\equiv$SiH did not 
capture electrons and holes. 

Our simulation shows that the Si-Si bonds in Si$_3$N$_4$ and SiO$_2$ can be 
both the electron and hole traps. We have obtained also that the $\equiv$Si* 
defect in Si$_3$N$_4$ and SiO$_2$ can capture an electron. However, the ESR 
experiments do not show the existence of this defect in Si$_3$N$_4$ and 
SiO$_2$ [12]. 

The recent density functional calculations of the $\equiv$Si* and 
$\equiv$ Si$_2$N* defects in silicon nitride [13] confirm that both defects 
can capture an electron. 

Thus, the calculations of the electronic structure of the main intrinsic 
defects in SiO$_x$N$_y$ in different charge states support the identification 
of the $\equiv$Si$_2$N* and $\equiv$SiO* defects with the electron traps. The 
hole traps are related with the Si-Si bonds. This conclusion is in agreement 
with the results of experiments [14]. It was suggested earlier by Robertson 
and Powell [15] that the negatively charged nitrogen defect $\equiv$Si$_2$N* 
can be a hole trap. Our results agree with it and show also that the electron 
localization by neutral defect $\equiv$Si$_2$N* leads to the ESR signal 
disappearance. This effect was experimentally observed earlier [16]. The 
results of calculations allow to explain the mechanism of creation and removal 
of the intrinsic defects and to understand the chemistry of the main 
technological steps in the gate silicon oxynitride fabrication. With these 
results the nature of the traps responsible for the electron and hole capture 
in gate oxynitride and so the main source of the silicon devices failure 
become more clear. 

\begin{center}
{\bf References}
\end{center}

$^*$ Electronic address: grits@isp.nsc.ru \\
1.  E.P. Gusev, H.C. Lu, E.L. Garfunkel, T. Gustafsson, and M.L. Green, 
    IBM J. Research and Develop. {\bf 43} (1999) 1. \\
2.  P.M. Lenahan and P.V. Dressendorfer, J. Appl. Phys. {\bf 55} (1984) 3495. 
    \\
3.  V.A. Gritsenko, J.B. Xu, Y.H. Ng, R.W.M. Kwok, and I.H.Wilson, 
    Phys. Rev. Lett. {\bf 81} (1998) 1054. \\
4.  I.A. Britov, V.A. Gritsenko, and Yu.N. Romaschenko, Sov. Phys. JETP 
    {\bf 62} (1985) 321. \\
5.  V.A. Gritsenko, Yu.N. Morokov, and Yu.N. Novikov, Appl. Surf. Sci. 
    {\bf 113/114} (1997) 417. \\
6.  Yu.N. Morokov, V.A. Gritsenko, Yu.N. Novikov, and H. Wong, 
    Microelectronic Engineering {\bf 48} (1999) 175. \\
7.  W.L. Warren, J. Robertson, and J. Kanicki,  Appl. Phys. Lett. {\bf 63} 
    (1993) 2685. \\
8.  L. Martin-Moreno, E. Martinez, J.A. Verges, and F. Yndurain, Phys. Rev. 
    {\bf B35} (1987) 9683. \\
9.  W.L. Warren, P.M. Lenahan, and S.E. Curry,  Phys. Rev. Lett. {\bf 65} 
    (1990) 207.\\
10.  J.T. Yount and P.M. Lenahan, J. Non-Cryst. Solids {\bf 164-166} (1993) 
    1069. \\
11. A.H. Edwars and W.B. Fowler, in {\it Structure and Bonding in 
    Noncrystalline Solids}, G.E. Walrafen and A.G. Revesz, Eds. 
    (Plenum Press, New York and London, 1986) 139-155. \\
12. V.A. Gritsenko and A. D. Milov, JETP Lett. {\bf 64} (1996) 531. \\
13. G. Pacchioni and D. Erbetta, Phys. Rev. {\bf B61} (2000) 15005. \\
14. V.A. Gritsenko, I.P. Petrenko, and S.N. Svitasheva, Appl. Phys. Lett. 
    {\bf 72} (1998) 462. \\
15. J. Robertson and M.J. Powell, Appl. Phys. Lett. {\bf 44} (1984) 415. \\
16. I.A. Chaiyasena, P.M. Lenahan, and G.J. Dunn, Appl. Phys. Lett. {\bf 58} 
    (1991) 2141. 

\end{document}